# Active Nonlocal Metasurfaces

Stephanie C. Malek[1], Adam C. Overvig[1,2], Sajan Shrestha[1], and Nanfang Yu[1,*]

[1]Department of Applied Physics and Applied Mathematics, Columbia University, New York, New York 10027, USA
[2]Photonics Initiative, Advanced Science Research Center at the Graduate Center of the City University of New York, New York, New York 10031, USA
[*]Correspondence to: ny2214@columbia.edu

**Abstract:** Actively tunable and reconfigurable wavefront shaping by optical metasurfaces poses a significant technical challenge often requiring unconventional materials engineering and nanofabrication. Most wavefront-shaping metasurfaces can be considered 'local' in that their operation depends on the responses of individual meta-units. In contrast, 'nonlocal' metasurfaces function based on the modes supported by many adjacent meta-units, resulting in sharp spectral features but typically no spatial control of the outgoing wavefront. Recently, nonlocal metasurfaces based on quasi-bound states in the continuum have been shown to produce designer wavefronts only across the narrow bandwidth of the supported Fano resonance. Here, we leverage the enhanced light-matter interactions associated with sharp Fano resonances to explore the active modulation of optical spectra and wavefronts by refractive index tuning and mechanical stretching. We experimentally demonstrate proof-of-principle thermo-optically tuned nonlocal metasurfaces made of silicon, and numerically demonstrate nonlocal metasurfaces that thermo-optically switch between distinct wavefront shapes. This meta-optics platform for thermally reconfigurable wavefront-shaping requires neither unusual materials and fabrication nor active control of individual meta-units.

## 1  Introduction

Metasurfaces are optically thin planar structured photonic devices [1,2]. We can consider metasurfaces to be local if the independent scattering events of individual meta-units dictates device behavior, or nonlocal if many adjacent meta-units support a collective mode that governs the response of the device [3]. Generically, local metasurfaces shape the wavefront across a broad spectrum, and may achieve functionalities such as lensing and holography. Nonlocal metasurfaces, in contrast, have sharp spectral control but typically without wavefront shaping capabilities, with prototypical examples including guided mode resonance filters [4] and photonic crystal slabs [5].

A significant ongoing challenge in metasurfaces is active tunability and reconfigurability of device functionality to realize devices such as varifocal metalenses or switchable metasurface holograms [6,7]. A few emerging mechanisms of active tunability in local metasurfaces include mechanical strain, thermal or electrical control with designer materials, and complex electrical actuation of individual meta-units. By modifying phase profiles via mechanical strain, local metasurfaces on stretchable substrates have yielded mechanically tunable zoom metalenses [8,9] and switchable metasurface holograms [10]. Electrical methods for actuating these stretchable metasurfaces have been introduced to simplify the mechanical system, including dielectric elastomeric actuators [11]. Phase change materials have been employed in local metasurfaces to realize thermally tunable devices such as a varifocal metalens with Ge-Sb-Se-Te meta-units [12] and dynamic meta-holograms based on vanadium dioxide [13]. Two-dimensional materials have also shown promise for active optics, including a tunable Fresnel zone plate based on the electrically tunable excitonic resonance in monolayer $WS_2$ [14]. More functionally versatile but electrically cumbersome approaches to tunable metasurfaces require electrical gating of individual meta-units, realizing, for example, one-dimensional beam steering by electrically control of liquid crystals [15] or InSb carrier concentration [16], and two-dimensional beam steering with a network of indium tin oxide-based electrodes [17].

In parallel, recent efforts towards active tuning of nonlocal metasurfaces have demonstrated tuning of the placement and linewidth of spectral features. Mechanical tuning of the resonant wavelength has been demonstrated in stretchable dielectric photonic crystals [18,19] and plasmonic lattices [20,21]. Other approaches include thermal [22] or electrical [23] tuning of photonic crystals. Our previous works computationally demonstrated tunable nonlocal metasurfaces based on quasi-bound states in the continuum (q-BICs), using electro-optic tuning of silicon to tune resonant frequencies [24] and mechanical tuning to control quality factors by active symmetry breaking [25]. Nonlocal metasurfaces have the notable advantage over local metasurfaces of enhanced light-matter interactions due to the long optical lifetime states, increasing the efficacy of minute refractive index changes. However, they have so far been limited to modulating optical spectra, with no active control over the shape of the optical wavefronts.

In our previous works, we have demonstrated theoretically [25,26] and experimentally [27] a generalization of nonlocal metasurfaces capable of shaping an optical wavefront by spatially varying the polarization properties of q-BICs. In particular, by using a geometric phase associated with the linear [26] or circular [28] dichroism of Fano resonances support by suitably perturbed photonic crystal slabs, nonlocal metasurfaces have been demonstrated that shape light only across the narrow spectral bandwidth of the resonance. This platform inherits both the enhanced light-matter interactions of the arbitrarily narrow linewidths of q-BICs [29] and the metasurface physics encapsulated by the Generalized Snell's Law [30].

In this work, we experimentally demonstrate thermo-optic tuning of resonant frequencies in silicon nonlocal metasurfaces based on q-BICs, a simple platform that can be extended to realize more complex nonlocal metasurfaces. We then explore the spatial and spectral reconfigurability of nonlocal metasurfaces. We show that wavefront-shaping of narrowband incident light can be turned on and off by shifting the resonant wavelength through refractive-index tuning (**Fig. 1A**), and that both wavefront shape and resonant wavelength can be tuned by mechanical strain. Our independently tunable metasurfaces can be cascaded to achieve reconfigurable or tunable multifunctional devices (**Fig. 1B**). Additionally, we demonstrate a multifunctional metasurface exhibiting two distinct functionalities while switching between its two resonances (**Fig. 1C**).

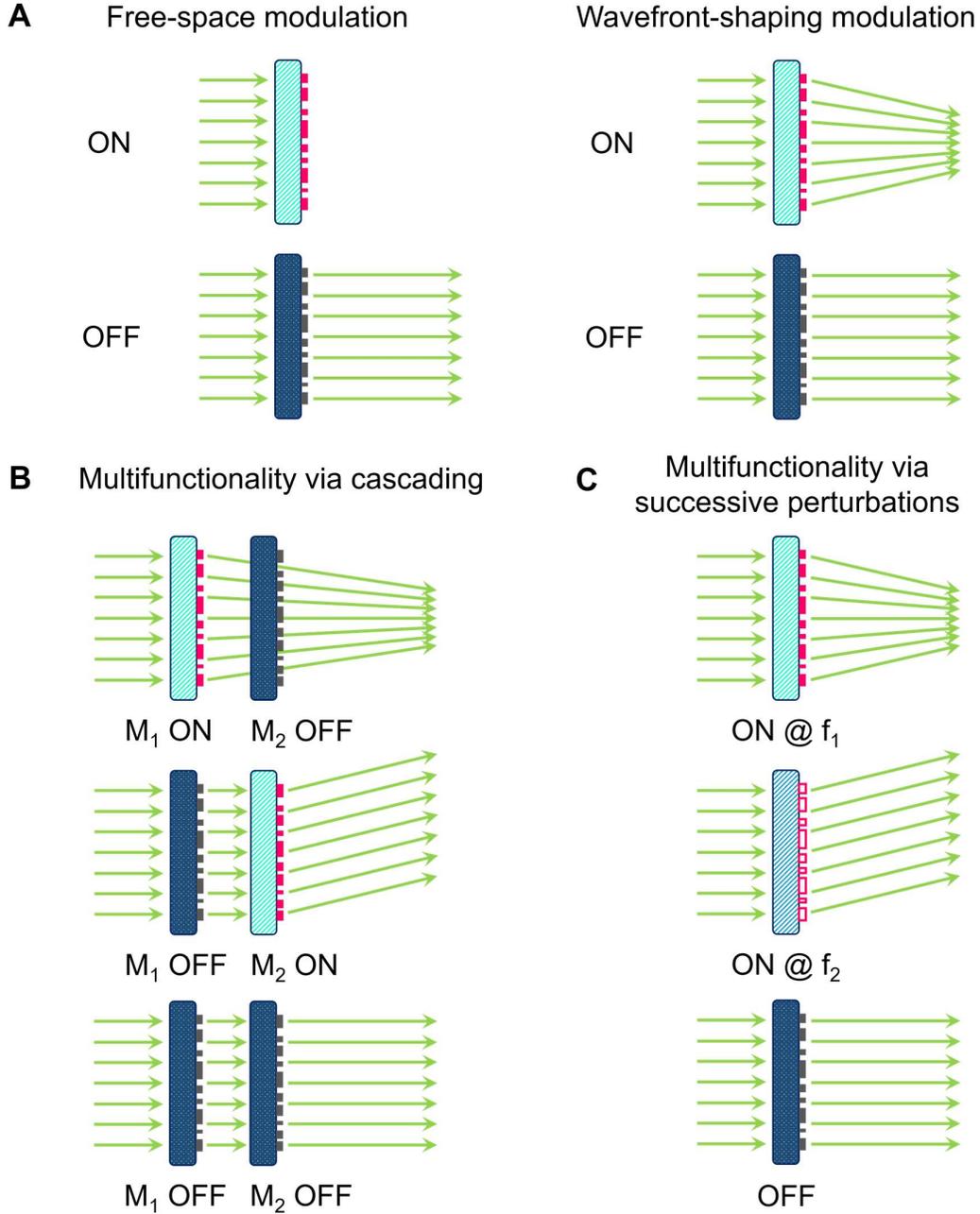

**Fig. 1**. Schematics illustrating several functionalities of nonlocal metasurface modulators. (A) Free-space modulation (left) and wavefront-shaping modulation (right) based on single nonlocal metasurfaces that can be switched on or off when its resonance is aligned with or detuned from the frequency of the incident light. (B) Multifunctional switchable modulation based on cascading nonlocal metasurfaces with distinct wavefront-shaping capabilities at distinct resonant frequencies. (C) Multifunctional switchable modulation based on a single metasurface containing orthogonal perturbations to implement distinct wavefront-shaping capabilities at distinct resonant frequencies.

# 2 Metasurfaces supporting quasi-bound states in the continuum

Bound states in the continuum (BICs) are states that are momentum-matched to free space but are nevertheless bound to a device [31]. BICs can be either accidental in that they incidentally have a coupling coefficient to free space of zero or symmetry protected in that a symmetry incompatibility forbids a mode from coupling to free space. We consider here symmetry-protected BICs that are controlled through a symmetry-breaking perturbation, yielding states (i.e., q-BICs) that radiate to free space with designer polarization dependence and quality factors (Q-factors), which vary inversely with the perturbation strength ($\delta$) as $Q\sim1/\delta^2$ [24,29]. We focus on period-doubling perturbations that double the period along a real-space dimension and therefore halve the period in k-space. This effectively folds the bandstructure such that modes that were previously bound and under the light line are folded to the $\Gamma$-point where they can be excited by free-space light at normal incidence. These q-BICs obey selection rules [25] governing whether excitation of a given mode is forbidden or allowed according to the symmetries of the mode, perturbation, and polarization of incoming light. Briefly, light at normal incidence will couple from free space to a dimerized structure only if the out-of-plane component of the mode's E or H field gives an effective net in-plane dipole moment in the direction of the incident polarization. The variety of possible optical responses has been cataloged based on symmetry degeneration from square or hexagonal lattices [24], a process which presents three key insights.

First is a rational design scheme for dimerized photonic crystal slabs with the following steps: (1) choose a real-space mode profile (e.g., large modal overlap with tunable materials) by choice of a photonic crystal structure and one of its high-symmetry modes, (2) adjust geometrical parameters to minimize photonic band curvature, as flat bands are associated with small group velocities and thus reduced device footprints, and (3) choose a type of dimerizing perturbation according to the selection rules [25] to target a specific free-space polarization state, and tune the perturbation strength to achieve desired Q-factors. The second key insight is that metasurfaces with local *p2* plane group symmetry imparts a geometric phase to incident circularly polarized light. The third insight is that we can add successive independent perturbations to a single nonlocal metasurface for realizing multifunctional device operation [24].

The first insight allows us to design free-space metasurface modulators with small footprints, desired Q-factors, and most importantly, deliberate field overlap with the active material. The second insight allows for wavefront-shaping resonant metasurfaces by tiling meta-units with distinct geometric phase responses. Taken together, these two insights uniquely enable a wavefront-shaping modulator that shapes the wavefront when narrowband incident light is aligned with the resonance, but leaves the wavefront shape unaltered when the incident light and resonance are intentionally misaligned by modulation. Finally, adding successive perturbations controlling two q-BICs with a small spectral separation allows us to demonstrate a multifunctional metasurface modulator that can be modulated to switch between distinct wavefront-shaping functionalities.

## 3 Free-space thermo-optic modulators

The rational design process described above can be leveraged to design free-space modulators based on metasurfaces that provide a large field overlap with the active material. We begin by developing thermo-optic nonlocal metasurface modulators in silicon and then extend this principle to wavefront-shaping nonlocal metasurfaces. Silicon is a common choice of active material with a thermo-optic coefficient of $\sim2\times10^{-4}$ K$^{-1}$ near telecommunications wavelengths [32]. The simplest case of a dimerized nonlocal metasurface is a one-dimensional (1D) grating. In 1D dimerizing perturbation we can either create a "gap perturbation" where the gap size between grating fingers alternates between two values for every other finger, or a "width perturbation" where the width of every other finger alternates between two values. In both cases, the perturbation doubles the period of the structure in real space, which halves the period in *k*-space. This effectively folds the first Brillouin zone, resulting in $0^{th}$-order diffractive modes excitable from free space (above the light line) in the perturbed structure that were bound modes (below the light line) in the unperturbed structure.

For a silicon metasurface on an insulator substrate we choose a gap perturbation as shown in **Fig. 2A** and consider the perturbation strength $\delta$ to be the difference in the widths of adjacent gaps. The structure supports a flat band (**Fig. 2B**) due to mode hybridization as a result of out-of-plane symmetry breaking by the substrate [33]. The electric-field mode profile (**Fig. 2A**) for excitation polarized parallel to the fingers has a good overlap with the silicon fingers. We fabricate these devices on a silicon-on-insulator substrate with electron beam lithography and dry etch with an alumina hard mask. We demonstrate a set of devices with a footprint of 500 μm × 500 μm and varying perturbation strengths (**Fig. 2C**). The latter control the optical lifetime with smaller perturbations producing higher measured Q-factors of up to Q~300 (**Fig. 2D**). Measurements of the reflection spectra of these devices over a 100°C temperature range show a 4.6-nm shift in the resonant wavelength and an extinction ratio of 2.4 at λ=1549 nm (**Fig. 2E**). We note that this device can also be used to enhance third harmonic generation in argon gas in the gaps between the silicon fingers due to enhanced optical fields there [34].

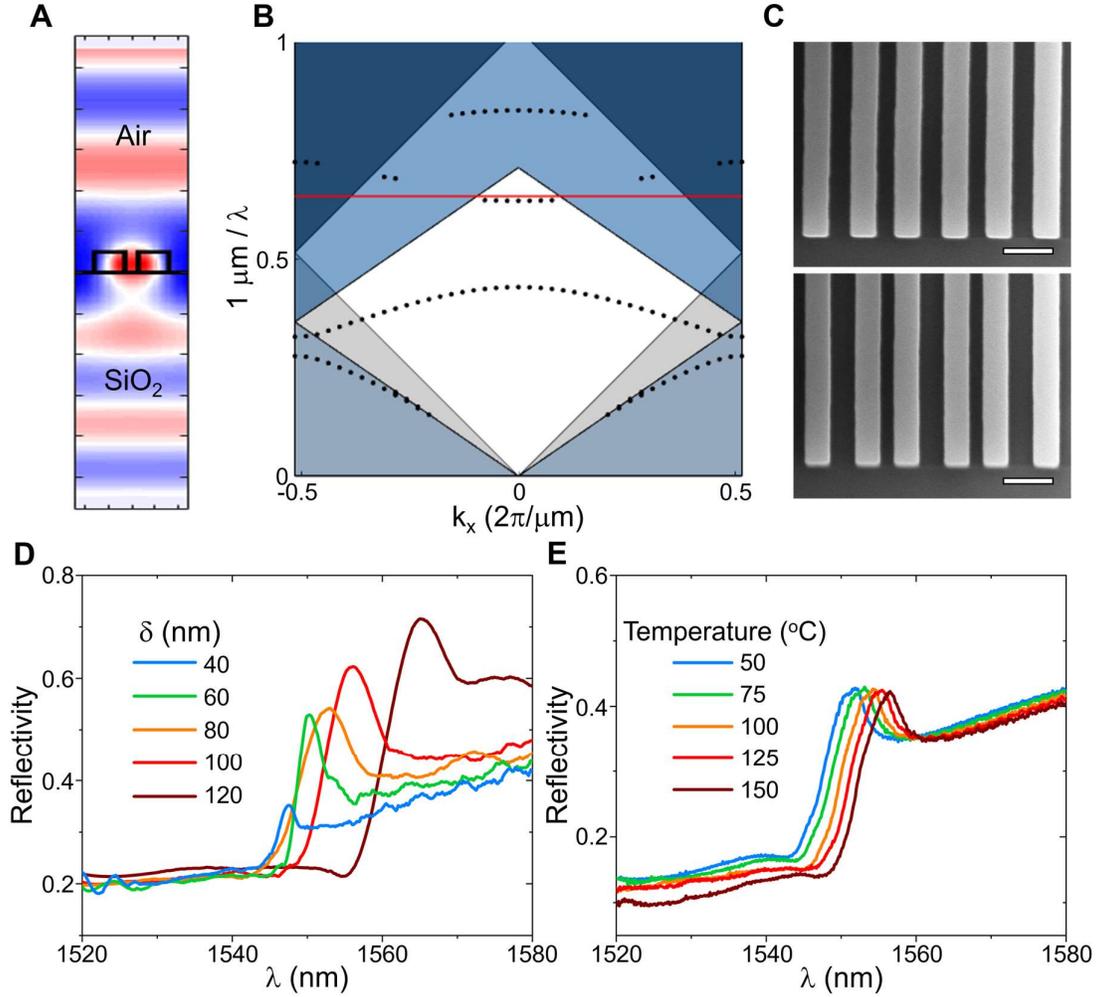

**Figure 2.** Design and experimental results of 1D free-space metasurface modulators. (A) Electric-field profile of a q-BIC mode at λ=1.55 μm over one period of a "gap perturbed" 1D metasurface. The cross-sections of silicon grating fingers are outlined in black and the incident polarization is along the fingers. Device dimensions: period of 950 nm, finger width of 270 nm, finger height of 250 nm, and gap perturbation of δ=20 nm (i.e., difference in size of adjacent gaps of 20 nm). (B) Band diagram of the 1D metasurface. Red line indicates λ=1.55 μm. (C) Scanning electron microscope (SEM) images of fabricated structures with δ=40 nm (top) and δ=100 nm (bottom). Scale bar: 500 nm. (D) Measured reflection spectra of devices with different δ. (E) Measured reflection spectra of a device with δ=60 nm at five different temperatures.

We also demonstrate polarization-insensitive thermo-optic modulators in silicon. According to the selection-rule catalog [25], polarization insensitive behavior requires degenerate E-type modes that are preserved by four-fold rotational symmetry. We choose a two-dimensional (2D) structure belonging to the *p4g* plane group as depicted in **Fig. 3A** and adjust the period and fill factor to minimize the band curvature for a mode at the telecommunications wavelengths (**Fig. 3B**). The out-of-plane electric-field profiles on resonance of the degenerate modes for x- and y-polarized incident light show a large modal overlap with the metasurface (**Fig. 3A**). We fabricate 2D devices with a range of perturbation strengths (**Fig. 2C**) and experimentally obtain Q-factors as high as Q~600 for a device consisting of rectangular silicon pillars with in-plane dimensions of 505 nm × 425 nm (i.e., δ=80 nm) (**Fig. 2D**). Q~600 is comparable to the highest experimentally realized Q-factors in q-BIC metasurfaces [35–37]. For a device with Q~290, measured reflection spectra show a 3.2-nm shift in resonant wavelengths over a 100°C temperature range and an extinction ratio of 1.18 at λ=1529 nm (**Fig. 3E**).

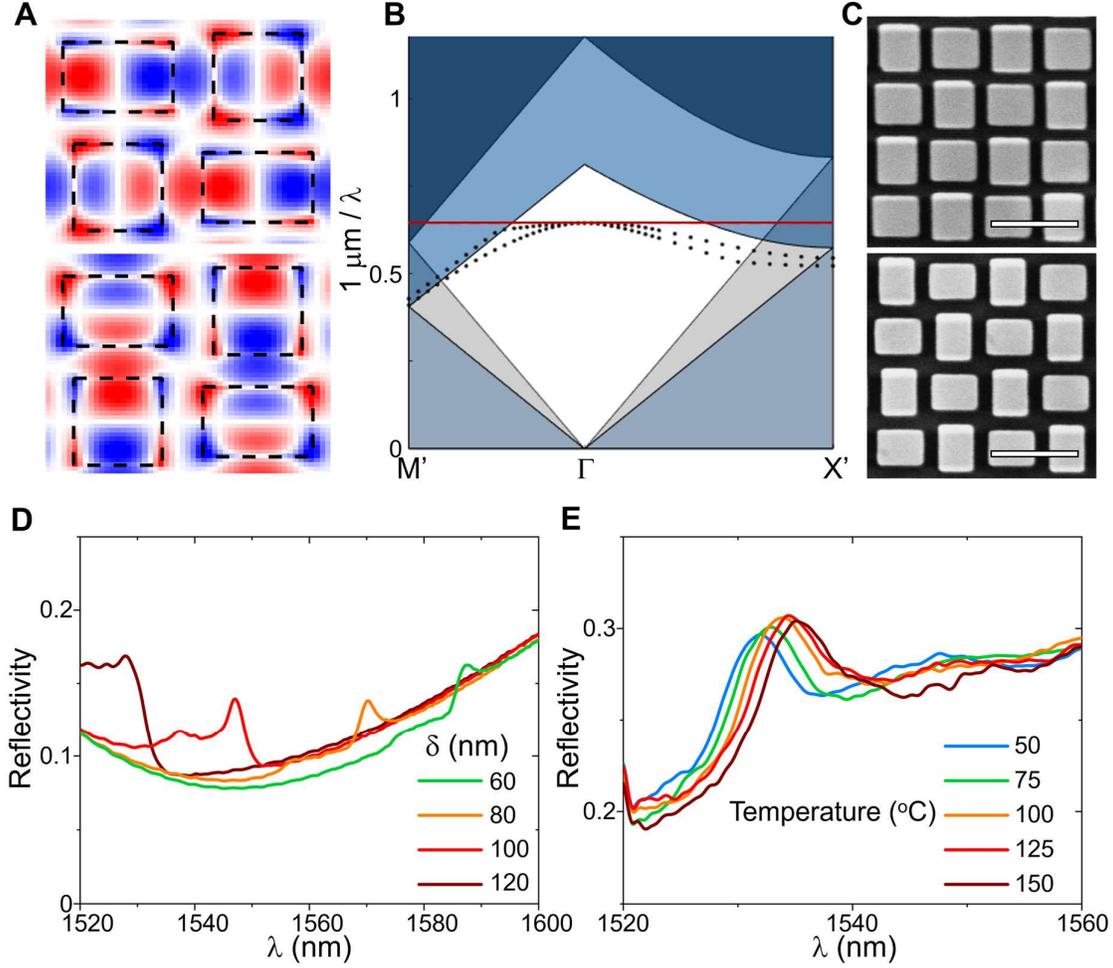

**Figure 3.** Design and experimental results of 2D polarization-insensitive metasurface modulators. (A) Mode profiles (i.e., out of plane electric-field component of degenerate E-modes) of a 2D structure belonging to the *p4g* plane group excited by y-polarized (top) and x-polarized (bottom) incident light. (B) Folded band diagram of an unperturbed 2D device consisting of a square lattice of square silicon pillars. Red line indicates λ=1.55 µm. Device dimensions: unperturbed period of 650 nm, square width of 450 nm, and silicon thickness of 250 nm. (C) SEM images of fabricated 2D devices with δ=60 nm (505 nm × 445 nm rectangles) (top) and δ=120 nm (505 nm × 385 nm rectangles) (bottom). Scale bar: 1 µm. (D) Measured reflection spectra of devices with different δ. (E) Measured reflection spectra at five different temperatures of a device with a period of 1300 nm and consisting of rectangular pillars with cross-sectional dimensions of 525 nm × 440 nm.

# 4 Wavefront-shaping modulators

Next, we design thermally tunable wavefront-shaping metasurfaces by choosing a device geometry that provides a large modal overlap with the active material and a spatially tailorable geometric phase [24,25] using the proof of principle platform demonstrated in Figures 2 and 3. Our selection-rule catalog [24] shows that in a meta-unit belonging to the *p2* plane group, for example, two rectangular apertures defined in a thin dielectric film (**Fig. 4A**), the in-plane rotation angle α of the apertures controls the far-field polarization angle ϕ that can couple to a q-BIC mode, such that ϕ~2α. This relationship between ϕ and α derives from the "parent-child" symmetry relation between higher symmetry plane groups *pmm* and *pmg* as "parent" groups and p2 as the "child" group. Comparing the geometry and selection rules of the *pmm* and *pmg* plane groups reveals that a change in-plane rotation angle α of 45° gives a 90° change in couplable polarization angle ϕ leading to the ϕ~2α relationship. This linear dichroism introduces a geometric phase twice as much as the conventional geometric phase of local metasurfaces: when circularly polarized light is incident onto the meta-unit, one factor of the geometric phase, $\Phi_{in}$=ϕ~2α, is produced from coupling into the linearly polarized q-BIC mode;

subsequently, another factor of the geometric phase, $\Phi_{out}=\phi\sim2\alpha$, is produced when light couples out into the free space and is decomposed into circularly polarized light of opposite handedness. As such, there is a total geometric phase of $\Phi=\Phi_{in}+\Phi_{out}\sim4\alpha$. However, there is no geometric phase for transmitted light with the same handedness of circular polarization as that of the incident light, because $\Phi_{in}=\phi$ and $\Phi_{out}=-\phi$ cancel each other. In other words, we get a geometric phase of $\Phi\sim4\alpha$ because dichroic elements impart a geometric phase of $\Phi\sim2\phi$ and these nonlocal metasurfaces have $\phi\sim2\alpha$. Typical local metasurfaces have a $\phi\sim\alpha$ and therefore give a geometric phase $\Phi\sim2\alpha$. It is also relevant that the maximum transmission efficiency of converted light on resonance is ¼ as this nonlocal metasurfaces is a four-port system with ½ of the incident light transmitted and ½ of the incident light reflected on resonance where at most ½ of the transmitted light has converted circular polarization.

We begin designing the wavefront-shaping modulator by choosing *p2* structure with rectangular apertures in silicon and selecting a q-BIC TM mode with $A_1$ symmetry as shown in the transverse and longitudinal cross-sections of the mode profile (**Figs. 4B** and **4C**) to ensure a large modal overlap with the active material, silicon, for efficient thermo-optic modulation. For this mode, there is a Lorentzian transmission peak with $Q\sim150$ for light of converted circular polarization and a dip for light of unconverted circular polarization (**Fig. 4D**). We then confirm that geometric phase follows the $\Phi\sim4\alpha$ relationship (**Fig. 4E**). Meta-units with different values of α and thus different phase responses can be tiled to form spatially varying phase profiles, creating devices such as lenses and beam deflectors. The resulting devices shape the wavefront only on resonance and only for transmitted light of converted handedness of circular polarization.

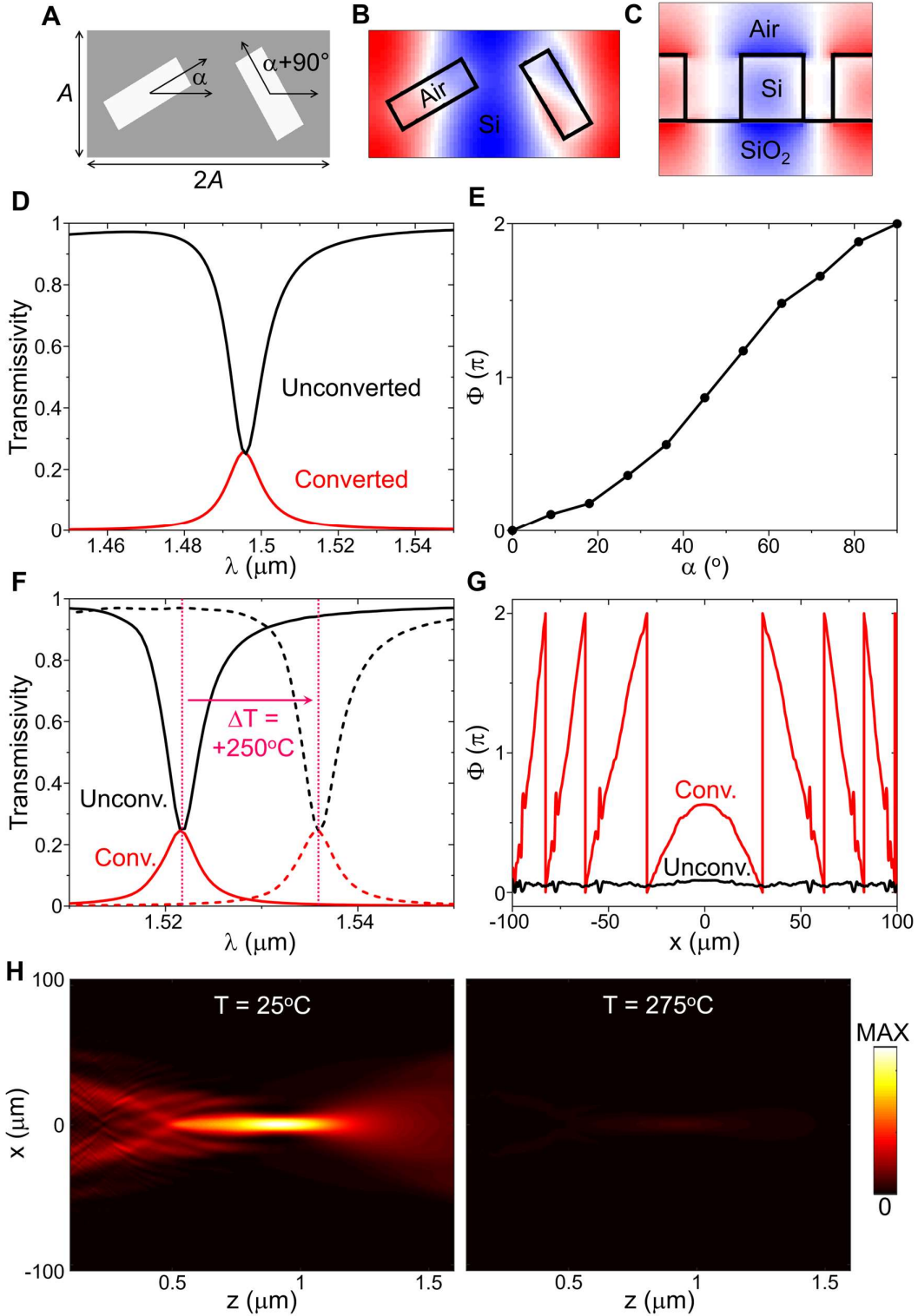

**Figure 4**. Design of a wavefront-shaping thermo-optic modulator in silicon. (A) Schematic of a meta-unit belonging to the *p2* plane group. It consists of rectangular apertures defined in a 250-nm thick silicon thin film. Here, A=400 nm, α is variable, and the aperture dimensions vary linearly with α between 100 x 275 nm at α=45° and 120 x 235 nm at α=0° or 90°. (B) Out-of-plane component of electric field ($E_z$) in a cross-section parallel to the substrate. (C) Distribution of $E_z$ in a cross-section perpendicular to the substrate. (D) Transmission spectra of one meta-unit (with periodic boundary condition) for light with converted circular polarization (red) and unconverted circular polarization (black) at 25°C. (E) Geometric phase, Φ, as a function of in-plane rotation angle α, showing a relation of Φ~4α. (F) Transmission spectra of light with converted circular polarization (red) and unconverted circular polarization (black) for a NA=0.1 nonlocal metalens at 25°C and 275°C. (G) Phase profile of the metalens

for light with converted circular polarization (red) and unconverted circular polarization (black). (H) Far-field intensity distributions at λ=1.521 µm of the metalens at 25°C (left) and 275°C (right).

With this meta-unit library, we create a cylindrical metalens with a numerical aperture (NA) of 0.1 and a dimension along the phase profile direction of 200 µm. Simulated transmission spectra of light with converted handedness of the device at 25°C (n=3.45) and 275°C (n=3.50) show a shift in the resonant wavelength of 14.0 nm and an extinction ratio of 37.9 at λ=1521 nm (**Fig. 4F**). Simulated far-field distributions of the metalens at λ=1521 nm (**Fig. 4H**) demonstrate that the device acts as a lens at 25°C but not at 275°C where little light of converted handedness is transmitted. Hence for narrowband incident light the device exhibits thermally switchable functionalities between that of a lens and that of an unpatterned substrate. We have previously demonstrated experimentally that nonlocal metasurfaces with distinct resonant wavelengths can be cascaded to achieve multifunctional behavior [27]. Cascaded and independently tunable nonlocal metasurface modulators could enable multifunctional switchable behavior by tuning the resonance of each metasurface to be aligned or misaligned with the narrowband incident light (**Fig. 1B**). We caution that aligning the resonances of multiple cascaded metasurfaces to the narrowband incident light does not provide a meaningful use case on account of its low efficiency as each metasurface can transmit only ~25% of the incident light on resonance for each circular polarization.

The pathway for switchable multifunctional thermo-optic modulators on a single metasurface requires successive orthogonal perturbations to control distinct q-BIC modes (**Fig. 1C**). We have previously proposed a scheme of successive orthogonal perturbations that produce independent geometric phases for up to four wavelengths: a single-layered metasurface can generate distinct wavefront shapes at spectrally separated resonances that are associated with orthogonal q-BIC modes [26]. Here, in order to leverage successive perturbations to design a thermally switchable multifunctional nonlocal metasurfaces we must consider one new design constraint: the spectral spacing between the adjacent orthogonal modes must be small enough for thermo-optic modulation to redshift the 'blue' resonance to align with the initial 'red' resonance. We satisfy this constraint by beginning with two modes that are degenerate in the unperturbed lattice, and apply a set of two perturbations simultaneously to lift the degeneracy and to supply the two modes with distinct selection rules [25]. **Figure 5A** shows a schematic of our chosen meta-unit consisting of apertures defined in a silicon thin film: gray circular apertures represent the unperturbed structure, and red and blue apertures represent the two orthogonal perturbations. Near telecommunications wavelengths, this meta-unit supports two TM modes (**Fig. 5B** right panels) each controlled by a separate perturbation such that rotating the red apertures controls the geometric phase of the redshifted mode following an approximately $\Phi_{red}=\alpha_{red}$ relation but has negligible impact on the blueshifted mode (**Fig. 5C** right panel). Conversely, the blue apertures impart a geometric phase $\Phi_{blue}\sim 4\alpha_{blue}$ to the blueshifted mode but not the redshifted one (**Fig. 5C** left panel). With this meta-unit library, we devise a device such that each of the orthogonal perturbations is tiled to create a distinct phase profile, leading to anomalous refraction of light to a distinct angle. A schematic superperiod of this device is shown in **Fig. 5D**. The far-field electric-field profiles at λ=1649 nm of light with converted circular polarization confirm that light is refracted to a 35-degree angle at 25°C and a 16.7-degree angle at 275°C (**Fig. 5E**). In this way, for narrowband incident light we can thermally switch the function of the nonlocal metasurface. We note that each of the resonances of the superperiod are spectrally blueshifted compared to the meta-unit design by the in-plane *k*-vector of the phase gradient [26]. Consequently, the spectral separation between the modes increases slightly compared to the meta-unit design therefore requiring a larger temperature differential to spectrally overlap the modes. This particular design considered only two distinct q-BIC resonances, but with more advanced optimization and potentially an active material affording stronger tunability, up to four distinct resonances and functionalities could be realized on a single metasurface [26].

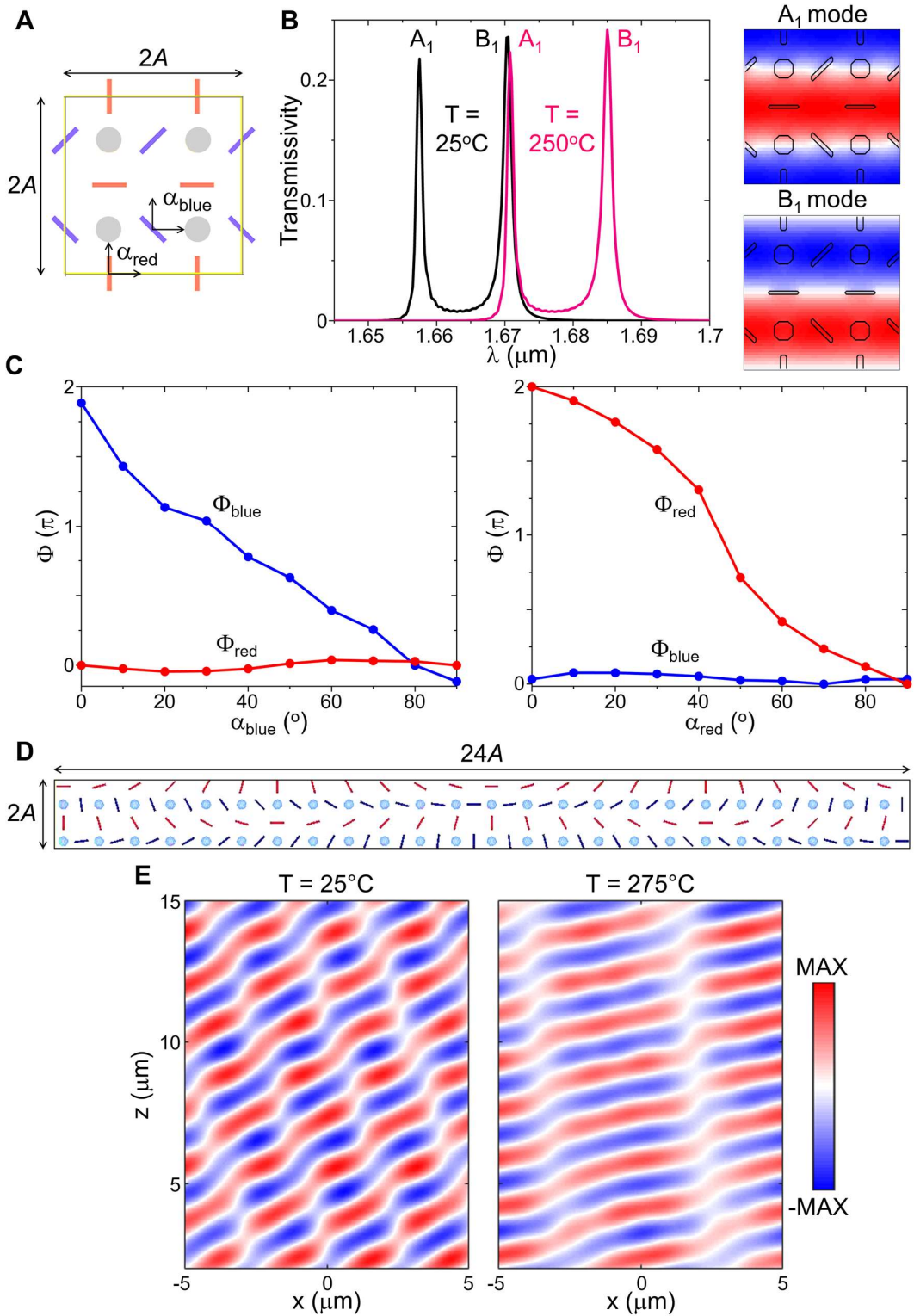

**Figure 5.** Design of a multifunctional wavefront-shaping thermo-optic modulator in silicon. (A) Composite meta-unit with two orthogonal perturbations (red and blue). Features denote apertures etched in a 250-nm thick silicon slab. The unperturbed lattice consists of a square array of circular apertures with a diameter of 125 nm and the lattice constant is A=900 nm. The colored rectangular apertures represent two types of perturbations and have a dimension of 175 nm × 25 nm. (B) Left panel: Transmission spectra of the meta-unit for light of converted circular polarization at 25°C (black curve) and 250°C (magenta curve). Right panels: Out-of-plane electric-field components of the 'blue' ($A_1$ type) and 'red' ($B_1$ type) modes. (C) Geometric phases control by the red

(right) and blue (left) apertures for the red and blue resonances, respectively. (D) One superperiod of a multifunctional beam deflector where red and blue apertures impart distinct phase gradients to the two modes. (E) Far field electric-field (real part) profiles at λ=1.649 µm of the multifunctional beam deflector at 25°C and 275°C.

Finally, we explore wavefront-shaping modulation in a mechanically stretchable system. Most commonly, local metasurfaces on stretchable substrates are stretched by uniform biaxial tensile stress to control the phase distribution and nonlocal metasurfaces are stretched by uniaxial tensile stress to control the quality factor or resonant wavelength. For our nonlocal metasurfaces, we consider nonuniform biaxial strain (**Fig. 6A**). We begin by devising a periodic 1D linear phase profile constructed with a *p2* meta-unit library with Q~150 and consisting of silicon pillars embedded in polydimethylsiloxane (PDMS), serving as a stretchable polymer substrate and superstrate. We then study the optical response of the embedded device as a function of stretching both along the phase gradient direction (x-direction) and orthogonal to it (y-direction), as shown in **Fig. 6B**. We exclude shear from our analysis and assume that the silicon pillars are simply displaced in the direction of the applied strain. In this case, *p2* symmetry is maintained upon stretching and the selection rules governing the geometric phase are unaffected. Stretching along the x-direction decreases the magnitude of the phase gradient and enables a beam steering functionality, and stretching in the y-direction alters the resonant frequency without affecting the phase gradient (in particular, stretching blueshifts the resonance frequency by lowering the effective refractive index). The resonant wavelength is a function of the phase gradient in our wavefront-shaping nonlocal metasurfaces [26]: the resonance redshifts with increased phase gradient for this mode. Additionally, deflection angle is dispersive with wavelength as governed by the generalized Snell's law [30], following the conventional dispersion of diffractive devices. However, by independently controlling the strain along both the x and y directions, simultaneous control of phase gradient and resonant frequency is possible.

With full-wave simulations, we calculate the resonant wavelength as a function of the period in the x and y directions (**Fig. 6C**) and mark contours of constant deflection angle calculated by the generalized Snell's law. We also map deflection angle as a function of period and mark contours of constant resonant wavelength (**Fig. 6D**). These results confirm the prediction of simultaneous control of resonant wavelength and deflection angle, and suggest two use cases for a stretchable nonlocal device: (1) a device that is stretched to deflect different wavelengths of light to a constant angle by following a contour lines in **Fig. 6C**, and (2) a device that is stretched to deflect the same wavelength of light to different angles by following a contour line in **Fig. 6D**. The first case enables devices with reconfigurable operating wavelength but without chromatic dispersion, while the latter case are active beam steering devices for specific operating wavelengths. More generally, these results demonstrate that judiciously stretched nonlocal metasurfaces based on flexible substrates may deflect light to a wide range of angles over a wide range of operating wavelengths. We have therefore demonstrated that nonlocal metasurface platforms may achieve simultaneous spectral and spatial control of light.

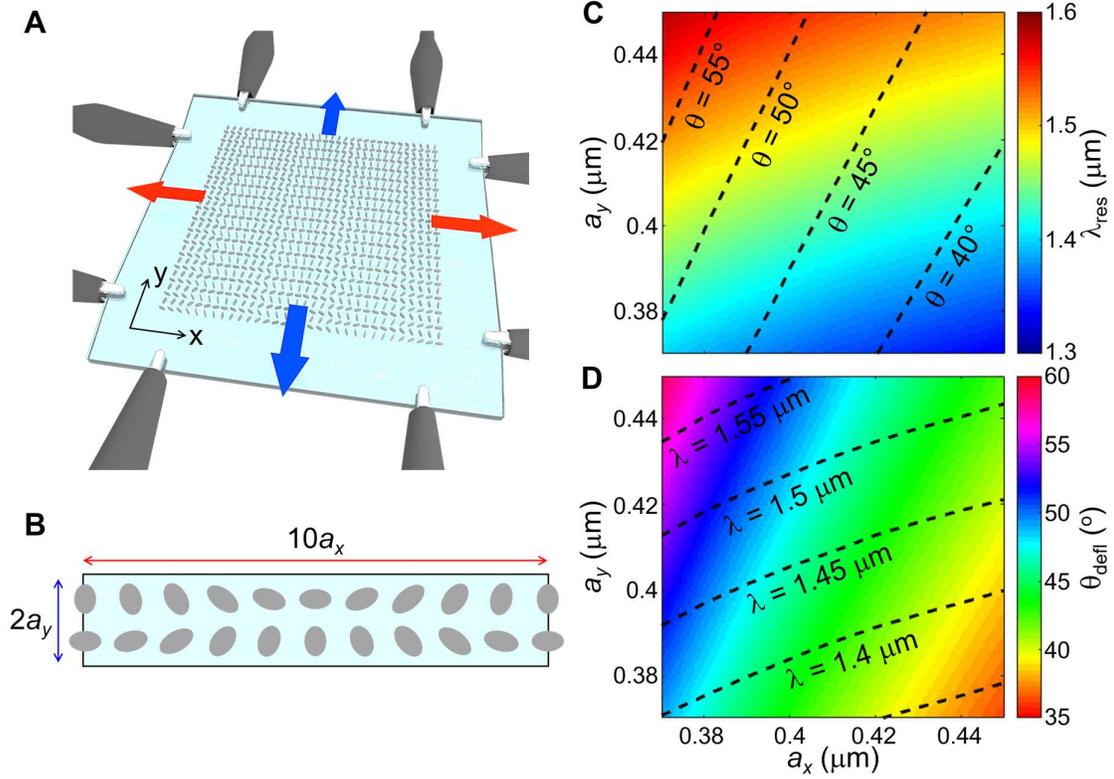

**Figure 6**. Design of mechanically tunable nonlocal metasurfaces. (A) Schematic of anisotropic stretching of a nonlocal metasurface. (B) Superperiod of a constant phase gradient metasurface with ellipse dimensions of height=300 nm, semi-major radius=160-183 nm, and semi-minor radius=100-110 nm. (C) Simulated map of resonant wavelength as a function of $a_x$ and $a_y$. (D) Simulated map of deflection angle as a function of $a_x$ and $a_y$.

# 5 Conclusion

We have experimentally demonstrated thermo-optic free-space modulators based on nonlocal metasurfaces and designed thermally and mechanically tunable wavefront-shaping nonlocal metasurfaces. Past this proof-of-principle demonstration, multifunctional behavior can be realized on a single metasurface incorporating orthogonal perturbations and cascaded tunable metasurfaces can achieve more complex multifunctionality. We note that these modulators behave differently than conventional tunable local and nonlocal metasurfaces both in the context of refractive index tuning (**Table 1**) and mechanical tuning (**Table 2**), and that the simultaneous control of spectra and wavefront afforded by nonlocal metasurfaces offers unique opportunities for active meta-optics. In particular, thermally tunable nonlocal metasurfaces provide a simpler platform compared with previous tunable local metasurface approaches by eliminating the need for designer materials (e.g., phase change materials), novel nanofabrication techniques, or electrical control of individual meta-units. Future work may incorporate electro-optic tuning or electrically-controlled thermo-optic tuning of nonlocal metasurfaces to provide a simple functional pathway for coveted electrically-reconfigurable optical metasurfaces modulating both the spectrum and shape of light.

13| Property | Local Geometric Phase Metasurface | Local Truncated Waveguide Metasurface | Local Huygens Metasurfaces [12,37,38] | Nonlocal Metasurface | Wavefront-Shaping Nonlocal Metasurface |
|---|---|---|---|---|---|
| Resonant Wavelength ($\lambda_{res}$) | N/A | N/A | Engineerable | Engineerable | Engineerable |
| Quality Factor (Q) | N/A | N/A | N/A | Trivial | Trivial |
| Phase Distribution ($\varphi(x,y)$) | Not Engineerable | Limited control (using material dispersion) | Engineerable | N/A | Engineerable (distinct $\varphi(x,y)$ at distinct resonances) |
| Band Structure | N/A | N/A | N/A | Trivial | Trivial |

**Table 1**: Comparison of properties tuned by refractive index tuning

| Property | Local Metasurface | Nonlocal Metasurface | Wavefront-Shaping Nonlocal Metasurface |
|---|---|---|---|
| Resonant Wavelength ($\lambda_{res}$) | N/A [8] or Trivial [9] | Engineerable [20,39] | Engineerable |
| Quality Factor (Q) | N/A [8] or Trivial [9] | Engineerable [24] | Engineerable |
| Phase Distribution ($\varphi(x,y)$) | Engineerable [8–10] | N/A | Engineerable |
| Band Structure | N/A | Engineerable | Engineerable |

**Table 2**: Comparison of properties tuned by mechanical strain

Last, we note that the primary limitations of our wavefront-shaping nonlocal metasurfaces are: (1) a maximum efficiency of ~25% that is associated with two instances of conversion of polarization states (i.e., circularly polarized incident light coupling into linearly polarized q-BIC modes, and coupling of the latter to circularly polarized output light), and (2) angular dispersion of the resonant wavelength (i.e., a relation between resonant frequency and the magnitude of phase gradient, which is dependent upon the flatness of the band structure), resulting in a tradeoff between numerical aperture and quality factor of devices [25,26]. The first challenge can be addressed by introducing chirality into nonlocal metasurfaces to achieve near unity efficiency in reflection mode [28], while the second is solvable by sophisticated bandstructure engineering or judicious adjustment of the local effective refractive index to maintain a constant resonant frequency across the device. Overcoming these challenges will enable high efficiency switchable meta-optics such as holograms and high-numerical-aperture lenses for use in display and imaging applications such as augmented reality.

**Acknowledgements**

The authors thank the Lipson Nanophotonics Group for assistance with optical measurements. The work was supported by the National Science Foundation (grant nos. ECCS-1307948 and QII-TAQS-1936359), the Defense Advanced Research Projects Agency (grant nos. D15AP00111 and HR0011-17-2-0017), and the Air Force Office of Scientific Research (grant no. FA9550-14-1-0389). A.C.O. acknowledges support from the NSF IGERT program (grant no. DGE-1069240). S.C.M acknowledges support from the NSF Graduate Research Fellowship Program (grant no. DGE-1644869).

# Bibliography

1.  N. Yu and F. Capasso, "Flat optics with designer metasurfaces," Nat Mater **13**, 139–150 (2014).
2.  P. Genevet, F. Capasso, F. Aieta, M. Khorasaninejad, and R. Devlin, "Recent advances in planar optics: from plasmonic to dielectric metasurfaces," Optica **4**, 139–152 (2017).
3.  H. Kwon, D. Sounas, A. Cordaro, A. Polman, and A. Alù, "Nonlocal Metasurfaces for Optical Signal Processing," Phys. Rev. Lett. **121**, 173004 (2018).
4.  S. Tibuleac and R. Magnusson, "Reflection and transmission guided-mode resonance filters," J. Opt. Soc. Am. A, JOSAA **14**, 1617–1626 (1997).
5.  S. G. Johnson, S. Fan, P. R. Villeneuve, J. D. Joannopoulos, and L. A. Kolodziejski, "Guided modes in photonic crystal slabs," Phys. Rev. B **60**, 5751–5758 (1999).
6.  I. Kim, G. Yoon, J. Jang, P. Genevet, K. T. Nam, and J. Rho, "Outfitting Next Generation Displays with Optical Metasurfaces," ACS Photonics **5**, 3876–3895 (2018).
7.  Q. He, S. Sun, and L. Zhou, "Tunable/Reconfigurable Metasurfaces: Physics and Applications," Research (2019).
8.  S. M. Kamali, E. Arbabi, A. Arbabi, Y. Horie, and A. Faraon, "Highly tunable elastic dielectric metasurface lenses," Laser & Photonics Reviews **10**, 1002–1008 (2016).
9.  H.-S. Ee and R. Agarwal, "Tunable Metasurface and Flat Optical Zoom Lens on a Stretchable Substrate," Nano Lett. (2016).
10. S. C. Malek, H.-S. Ee, and R. Agarwal, "Strain Multiplexed Metasurface Holograms on a Stretchable Substrate," Nano Lett. **17**, 3641–3645 (2017).
11. A. She, S. Zhang, S. Shian, D. R. Clarke, and F. Capasso, "Adaptive metalenses with simultaneous electrical control of focal length, astigmatism, and shift," Science Advances **4**, eaap9957 (2018).
12. M. Y. Shalaginov, S. An, Y. Zhang, F. Yang, P. Su, V. Liberman, J. B. Chou, C. M. Roberts, M. Kang, C. Rios, Q. Du, C. Fowler, A. Agarwal, K. Richardson, C. Rivero-Baleine, H. Zhang, J. Hu, and T. Gu, "Reconfigurable all-dielectric metalens with diffraction limited performance," arXiv:1911.12970 [physics] (2019).
13. X. Liu, Q. Wang, X. Zhang, H. Li, Q. Xu, Y. Xu, X. Chen, S. Li, M. Liu, Z. Tian, C. Zhang, C. Zou, J. Han, and W. Zhang, "Thermally Dependent Dynamic Meta-Holography Using a Vanadium Dioxide Integrated Metasurface," Advanced Optical Materials **7**, 1900175 (2019).
14. J. van de Groep, J.-H. Song, U. Celano, Q. Li, P. G. Kik, and M. L. Brongersma, "Exciton resonance tuning of an atomically thin lens," Nature Photonics 1–5 (2020).
15. A. I. Kuznetsov, A. E. Miroshnichenko, M. L. Brongersma, Y. S. Kivshar, and B. Luk'yanchuk, "Optically resonant dielectric nanostructures," Science **354**, (2016).
16. P. P. Iyer, M. Pendharkar, and J. A. Schuller, "Electrically Reconfigurable Metasurfaces Using Heterojunction Resonators," Advanced Optical Materials **4**, 1582–1588 (2016).
17. G. K. Shirmanesh, R. Sokhoyan, P. C. Wu, and H. A. Atwater, "Electro-optically Tunable Multifunctional Metasurfaces," ACS Nano (2020).
18. C. L. Yu, H. Kim, N. de Leon, I. W. Frank, J. T. Robinson, M. McCutcheon, M. Liu, M. D. Lukin, M. Loncar, and H. Park, "Stretchable Photonic Crystal Cavity with Wide Frequency Tunability," Nano Lett. **13**, 248–252 (2013).
19. M. L. Tseng, J. Yang, M. Semmlinger, C. Zhang, P. Nordlander, and N. J. Halas, "Two-Dimensional Active Tuning of an Aluminum Plasmonic Array for Full-Spectrum Response," Nano Lett. **17**, 6034–6039 (2017).
20. Y. Cui, J. Zhou, V. A. Tamma, and W. Park, "Dynamic Tuning and Symmetry Lowering of Fano Resonance in Plasmonic Nanostructure," ACS Nano **6**, 2385–2393 (2012).
21. T. Lewi, N. A. Butakov, H. A. Evans, M. W. Knight, P. P. Iyer, D. Higgs, H. Chorsi, J. Trastoy, J. D. V. Granda, I. Valmianski, C. Urban, Y. Kalcheim, P. Y. Wang, P. W. C. Hon, I. K. Schuller, and J. A. Schuller, "Thermally Reconfigurable Meta-Optics," IEEE Photonics Journal **11**, 1–16 (2019).
22. C. Qiu, J. Chen, Y. Xia, and Q. Xu, "Active dielectric antenna on chip for spatial light modulation," Scientific Reports **2**, 855 (2012).
23. A. C. Overvig, S. Shrestha, and N. Yu, "Dimerized high contrast gratings," Nanophotonics **7**, 1157–1168 (2018).
24. A. C. Overvig, S. C. Malek, M. J. Carter, S. Shrestha, and N. Yu, "Selection rules for quasibound states in the continuum," Phys. Rev. B **102**, 035434 (2020).
25. A. C. Overvig, S. C. Malek, and N. Yu, "Multifunctional Nonlocal Metasurfaces," Phys. Rev. Lett. **125**, 017402 (2020).
26. S. C. Malek, A. C. Overvig, S. Shrestha, and N. Yu, "Resonant Wavefront-Shaping Metasurfaces Based on Quasi-Bound States in the Continuum," in *Conference on Lasers and Electro-Optics* (2020).
27. A. Overvig, N. Yu, and A. Alu, "Chiral Quasi-Bound States in the Continuum," arXiv:2006.05484 [physics] (2020).
28. K. Koshelev, S. Lepeshov, M. Liu, A. Bogdanov, and Y. Kivshar, "Asymmetric Metasurfaces with High-Q Resonances Governed by Bound States in the Continuum," Phys. Rev. Lett. **121**, 193903 (2018).
29. N. Yu, P. Genevet, M. A. Kats, F. Aieta, J.-P. Tetienne, F. Capasso, and Z. Gaburro, "Light Propagation with Phase Discontinuities: Generalized Laws of Reflection and Refraction," Science **334**, 333–337 (2011).

**15**

30. C. W. Hsu, B. Zhen, A. D. Stone, J. D. Joannopoulos, and M. Soljačić, "Bound states in the continuum," Nature Reviews Materials **1**, 1–13 (2016).
31. G. Cocorullo, F. G. Della Corte, and I. Rendina, "Temperature dependence of the thermo-optic coefficient in crystalline silicon between room temperature and 550 K at the wavelength of 1523 nm," Appl. Phys. Lett. **74**, 3338–3340 (1999).
32. S. Cueff, M. S. R. Huang, D. Li, X. Letartre, R. Zia, P. Viktorovitch, and H. S. Nguyen, "Tailoring the Local Density of Optical States and directionality of light emission by symmetry-breaking," arXiv:1810.04034 [physics] (2018).
33. J. S. Ginsberg, A. C. Overvig, M. M. Jadidi, S. Malek, G. Patwardhan, G. Patwardhan, N. Swenson, N. Yu, and A. L. Gaeta, "Enhancement of harmonic generation in gases using an all-dielectric metasurface," in *Conference on Lasers and Electro-Optics (2019), Paper FM4M.7* (Optical Society of America, 2019).
34. S. Campione, S. Liu, L. I. Basilio, L. K. Warne, W. L. Langston, T. S. Luk, J. R. Wendt, J. L. Reno, G. A. Keeler, I. Brener, and M. B. Sinclair, "Broken Symmetry Dielectric Resonators for High Quality Factor Fano Metasurfaces," ACS Photonics **3**, 2362–2367 (2016).
35. F. Yesilkoy, E. R. Arvelo, Y. Jahani, M. Liu, A. Tittl, V. Cevher, Y. Kivshar, and H. Altug, "Ultrasensitive hyperspectral imaging and biodetection enabled by dielectric metasurfaces," Nature Photonics **13**, 390–396 (2019).
36. Y. Yang, I. I. Kravchenko, D. P. Briggs, and J. Valentine, "All-dielectric metasurface analogue of electromagnetically induced transparency," Nature Communications **5**, 5753 (2014).
37. A. Afridi, J. Canet-Ferrer, L. Philippet, J. Osmond, P. Berto, and R. Quidant, "Electrically Driven Varifocal Silicon Metalens," ACS Photonics **5**, 4497–4503 (2018).
38. P. P. Iyer, R. A. DeCrescent, T. Lewi, N. Antonellis, and J. A. Schuller, "Uniform Thermo-Optic Tunability of Dielectric Metalenses," Phys. Rev. Applied **10**, 044029 (2018).
39. J.-H. Choi, Y.-S. No, J.-P. So, J. M. Lee, K.-H. Kim, M.-S. Hwang, S.-H. Kwon, and H.-G. Park, "A high-resolution strain-gauge nanolaser," Nature Communications **7**, 11569 (2016).